\documentclass[12pt]{iopart}

\usepackage{graphicx}
\usepackage{dcolumn}
\usepackage{subfigure}
\usepackage{bm}

\newcommand{\bmath}{\begin{displaymath}}
\newcommand{\emath}{\end{displaymath}}

\newcommand{\be}{\begin{equation}}
\newcommand{\ee}{\end{equation}}
\newcommand{\bea}{\begin{eqnarray}}
\newcommand{\eea}{\end{eqnarray}}

\newcommand{\bmultl}{\begin{multline}}
\newcommand{\emultl}{\end{multline}}

\newcommand{\bsubeq}{\begin{subequations}}
\newcommand{\esubeq}{\end{subequations}}
\newcommand{\bitemize}{\begin{itemize}}
\newcommand{\eitemize}{\end{itemize}}

\newcommand{\re}{\mathrm{Re}}
\newcommand{\im}{\mathrm{Im}}
\newcommand{\aop}{\hat{a}}

\newcommand{\bmx}{\begin{bmatrix}}
\newcommand{\emx}{\end{bmatrix}}
\newcommand{\bsmx}{\begin{smallmatrix}}
\newcommand{\esmx}{\end{smallmatrix}}

\newcommand{\Eq}[1]{Eq.~(\ref{#1})}

\newcommand{\Sec}[1]{Section~\ref{#1}}
\newcommand{\Fig}[1]{Fig.~\ref{#1}}

\begin{document}

\title[Excitations of optically driven atomic condensate in a cavity]{Excitations of optically driven atomic condensate in a cavity: theory of photodetection measurements}

\author{Bar\i\c{s} \"{O}ztop$^1$, Mykola Bordyuh$^2$, \"{O}zg\"{u}r E M\"{u}stecapl{\i}o\u{g}lu$^3$ and Hakan E T\"{u}reci$^{2,1}$}
\address{$^1$ Institute for Quantum Electronics, ETH-Z\"urich, CH-8093 Z\"urich,
Switzerland}
\address{$^2$ Department of Electrical Engineering, Princeton University, Princeton, New Jersey 08544, USA}
\address{$^3$ Department of Physics, Ko\c{c} University, \.Istanbul, 34450, Turkey}
\ead{oztop@phys.ethz.ch}

\date{\today}

\begin{abstract}
Recent experiments have demonstrated an open system realization of the Dicke quantum phase transition in the motional degrees of freedom of an optically driven Bose-Einstein condensate in a cavity. Relevant collective excitations of this light-matter system are polaritonic in nature, allowing access to the quantum critical behavior of the Dicke model through light leaking out of the cavity. This opens the path to using photodetection based quantum optical techniques to study the dynamics and excitations of this elementary quantum critical system. We first discuss the photon flux observed at the cavity face and find that it displays a different scaling law near criticality than that obtained from the mean field theory for the equivalent closed system. Next, we study the second order correlation measurements of photons leaking out of the cavity. Finally, we discuss a modulation technique that directly captures the softening of polaritonic excitations. Our analysis takes into account the effect of the finite size of the system which may result in an effective symmetry breaking term.
\end{abstract}

\submitto{\NJP}

\maketitle


\section{Introduction}
In the study of strongly correlated systems and collective phenomena, light has traditionally assumed the role of a spectroscopic probe. Recent progress in the control of light-matter interactions through cavity QED has brought forth new systems where light and matter play equally important roles in emergent phenomena. Such hybrid light-matter systems are characterized by the existence of well-defined quasi-particles, polaritons, which are partly light partly matter-like.
A crucial feature of these systems is that they are inherently out of equilibrium due to unavoidable photon-leakage, giving rise to open system analogues of certain well-studied quantum many body Hamiltonians. Such systems may for instance be formed by scaling up standard single-cavity QED systems to lattices of cavities. Recent theoretical work in Cavity QED lattices has addressed the realization of superfluid-Mott insulator transition of polaritons~\cite{plenio06,hollenberg06,bose07,imamoglu10}, fractional quantum Hall states~\cite{bose08}, the Tonks-Girardeau gas in one-dimensional geometries~\cite{demler08,imamoglu09} and effective spin models~\cite{bose07,plenio07}  .

A different approach is to couple non-trivial states of matter to cavities, where the formation of polaritonic collective excitations may result in new emergent phenomena. One particular approach is to couple a Bose-Einstein Condensate (BEC) to a single mode of a high-finesse optical cavity. This results in tunable, long-range forces between atoms of the BEC that is mediated by the cavity field. A phase transition from a uniform BEC to a self-organized, density-modulated phase has been predicted~\cite{nagy08} as a function of the power of the laser driving the atoms transverse to the cavity axis and experimentally observed recently~\cite{esslinger10}. A similar self-organization transition has also been theoretically predicted~\cite{dr02,dr05} and observed for a thermal cloud of atoms~\cite{vuletic03}. 

Here we study the excitations of the zero-temperature BEC self-organization transition, which, in the experimentally realized parameter regime \cite{esslinger10}, was shown to be a faithful open-system realization of the single-mode Dicke quantum phase transition where spins correspond to the collective motional degrees of freedom of atoms. The transition is driven by softening of a polaritonic excitation as the critical point is approached. This provides access to the internal excitation dynamics close to the criticality through photons that leak out of the cavity mirrors. Thus well-established quantum optical measurement schemes appear as specially suited to monitor the intra-cavity excitation dynamics as well as the phase diagram. We first discuss the critical behavior of photon flux measured at the cavity mirror. We then study the signature of excitations in the second order correlation functions of leaking photons, a measurement that can readily be performed using standard quantum optical schemes such as the Hanbury-Brown-Twiss setup~\cite{hbt}. Finally, we consider a modulation technique that directly captures the softening of the relevant polaritonic mode through the photodetection of leaking photons. Our discussion will take into account the role of the finite size of the system that is relevant for a realistic experimental setting which was found to result in an effective symmetry-breaking term \cite{esslinger11}.

The paper is organized as follows. In~\Sec{sec:system} we summarize the system and the governing Hamiltonian together with mapping to the Dicke model. We then investigate the collective light-matter excitations in \Sec{sec:fluct}. Finally we discuss three different photodetection-based measurement schemes in \Sec{sec:measurement}.

\section{Model of the system} \label{sec:system}

We consider a Bose-Einstein condensate (BEC) of length $D$ consisting of atoms with an internal optical transition $\omega_{a}$. This atomic condensate is coupled to a single mode of frequency $\omega_c$ and decay rate $\kappa$ of a high finesse cavity of length $L$, and is driven by a laser with frequency $\omega_p$ from a direction perpendicular to the cavity axis~\cite{esslinger10}. In the dispersive regime of driving relevant to recent experiments ($|\Delta_a| = |\omega_p - \omega_a| \gg \gamma_{a}$, where $\gamma_{a}$ is the atomic linewidth), the minimal model that captures the essential features of the system is (see~\ref{sec:app-1} for details):
\bea
\hat{H} &=& -\hbar\Delta_c \hat{a}^{\dagger} \hat{a} + \int dx \, \hat{\Psi}^{\dagger}(x) \left[ \frac{-\hbar^2}{2m}\frac{d^2}{dx^2} + V(\bm{x}) \right. \nonumber \\
&+& \left. \hbar U_0 |\varphi_c (x)|^2 \hat{a}^{\dagger}\hat{a} + \hbar\eta \varphi_c (x) (\hat{a}^{\dagger} + \hat{a}) \right] \hat{\Psi}(x),
\label{eq:manybody_H}
\eea
where $U_0 = g_0^2/\Delta_a$ and $\eta = \Omega_p g_0/\Delta_a$. $\Delta_c$ is the pump-cavity detuning, $\Omega_p$ is the pump Rabi frequency, $g_0$ is the atom-cavity coupling strength and $\varphi_c (x)$ is the cavity mode function. We will neglect the contact interactions between atoms of the condensate, which is not essential for the physics discussed here. It is also possible to map the Hamiltonian in Eq.~(\ref{eq:manybody_H}) to a Bose-Hubbard model in the limit of strong particle-particle interactions to discuss incompressible Mott-insulator phases~\cite{vidal-pra10}.

As a function of the tunable pump power $\eta$, this model displays a phase transition from a homogenous condensate to a density-modulated phase~\cite{nagy08}. Below a threshold power $\eta_c$, the intracavity (mean) field is vanishingly small and the cloud is in the ground state of only the external trapping potential (we assume $T=0$). As the critical point is crossed, the atoms self-organize into a crystalline order. This in turn results in a non-zero cavity mean-field through the Bragg-scattering of the pump photons from the density-modulated atomic cloud into the cavity. In the organized phase, the system chooses spontaneously between two energetically identical density-modulated configurations that are shifted by half a pump wavelength. This self-organization phase transition was observed experimentally for thermal atomic gas~\cite{vuletic03} and for a cloud of atomic BEC~\cite{esslinger10}.

For the experimental conditions of Ref.~\cite{esslinger10}, the self-organization transition can also be seen as the Dicke model phase transition from a normal to a superradiant phase~\cite{esslinger10,nagy10}. We will confine ourselves to this regime but take steps to accurately model the experimental conditions. We assume a BEC considerably smaller than the cavity size, imposed by the external trapping potential $V(x)$ and expand the field operator
\be
\hat{\Psi}(x) = \sum_n \hat{c}_n \phi_n (x)
\label{eq:Psi_expand}
\ee
Here, $\phi_n (x)$ is the atomic single-particle basis satisfying $\left[ \frac{-\hbar^2}{2m}\frac{d^2}{dx^2} + V(x) \right] \phi_n (x) = \omega_n \phi_n(x)$. We will assume $V(x)$ to impose Neumann Boundary conditions at $x = (L - D)/2 \pm d$ and $x = (L + D)/2 \pm d$, allowing for an asymmetric placement of the trap by a length $d$ with respect to the cavity walls (we will assume $d/D \ll 1$). Then, $\phi_0 (x) = 1/\sqrt{D}$ is the uniform mode of the condensate with zero momentum. In the expansion~(\ref{eq:Psi_expand}), we keep only one additional mode having a relatively large overlap with cavity mode $\varphi_c(x)$, say $\phi_n (x)$, which becomes the dominant contribution in the Hamiltonian of Eq.~(\ref{eq:manybody_H}) when the cavity mean-field is non-zero. The corresponding wave-vector is $k_n \approx G$. Here it is assumed that these two modes satisfy the relation $\hat{c}_0^{\dagger}\hat{c}_0 + \hat{c}_n^{\dagger}\hat{c}_n = N$. Introducing the Schwinger representation for 'spins' $\hat{J}_{-} = \hat{c}_0^{\dagger}\hat{c}_n$, $\hat{J}_{+} = \hat{J}_{-}^{\dagger}$ and $\hat{J}_{z} = (\hat{c}_n^{\dagger}\hat{c}_n - \hat{c}_0^{\dagger}\hat{c}_0)/2$, the final Hamiltonian can be written as
\bea
\label{eq:dicke_H}
\hat{H}_D /\hbar&=& \omega \hat{a}^{\dagger}\hat{a} + \omega_0 \hat{J}_z + \frac{\lambda}{\sqrt{N}}(\hat{a}^{\dagger} + \hat{a})(\hat{J}_{+} + \hat{J}_{-}) \nonumber \\
&+& \frac{\lambda^\prime}{\sqrt{N}} (\hat{a}^{\dagger} + \hat{a}) \left(\frac{N}{2} - \hat{J}_z \right)
\eea
where $\omega = -\Delta_C + (NU_0 /D) \int dx |\varphi_c (x)|^2$, $\omega_0 = \omega_R = \hbar G^2 / 2m$, $\lambda = (\sqrt{N/D}) \eta \int dx \varphi_c(x) \phi_n (x) $ and $\lambda^\prime = (\sqrt{N/D}) \eta \int dx \varphi_c (x) \phi_0(x)$. We observe that a non-zero $\lambda^\prime$ acts as a symmetry-breaking bias field where its sign determines which of the two configurations of the condensate is preferred in the superradiant phase. Note that $\phi_0(x)$ is uniform, so the integral defining $\lambda'$ vanishes if the condensate is placed symmetrically with respect to the cavity mode function $\varphi_{c}(x)$. This was pointed out in recent experiments studying the process of symmetry breaking in real time through an interferometric heterodyne detection scheme~\cite{esslinger11}. Since we are interested in the critical behavior, an additional dispersive shift term~\cite{esslinger10} which is negligible near critical point is dropped in Eq.~(\ref{eq:dicke_H}).

For $\lambda' = 0$, Eq.~\ref{eq:dicke_H} is the well-known single-mode Dicke Hamiltonian. In the thermodynamic limit of $N\gg 1$, the Dicke Hamiltonian~(\ref{eq:dicke_H}) exhibits a quantum phase transition at a critical coupling strength $\lambda_c = (1/2)\sqrt{\omega \omega_0}$ from a normal phase with $\langle a \rangle=\langle J_-\rangle=0$  to a superradiant phase~\cite{lieb73-1,lieb73-2,hioe73-1,hioe73-2,walls73,duncan74,brandes-prl03,brandes-pre03} with $\langle a \rangle \neq 0$, $\langle J_- \rangle \neq 0$, breaking the parity symmetry of the Dicke Hamiltonian~(\ref{eq:dicke_H}) under $(\hat{a} \rightarrow -\hat{a}$, $\hat{J}_{-} \rightarrow -\hat{J}_{-})$. 

Finally, cavity losses at a rate $\kappa$ can be included through a Lindblad master equation approach~\cite{carmichael07}. This leads to an open-system analogue of the Dicke superradiance transition, which, most significantly results in a shift of the critical point to $\lambda_c = (1/2)\sqrt{(\omega_0 /\omega ) (\kappa^2 + \omega^2 )}$. A second important consequence is that the excitation spectrum and the dynamics becomes dissipative. The details of the mean-field analysis of the open-system Dicke model with and without the symmetry breaking field $\lambda^\prime$ is given in~\ref{sec:app-2} by introducing mean fields $\alpha = \langle \hat{a}\rangle $, $\beta = \langle \hat{J}_{-}\rangle$ and $w=\langle \hat{J}_{z}\rangle$. We note that mean-field approximation neglects the entanglement between the atomic and photonic subsystems which can have significant impact on the transient dynamics towards the steady state \cite{maschler-optcomm07}.

\section{Polaritonic excitations} \label{sec:fluct}

\begin{figure}[ht]
\centering
\includegraphics[width=16cm]{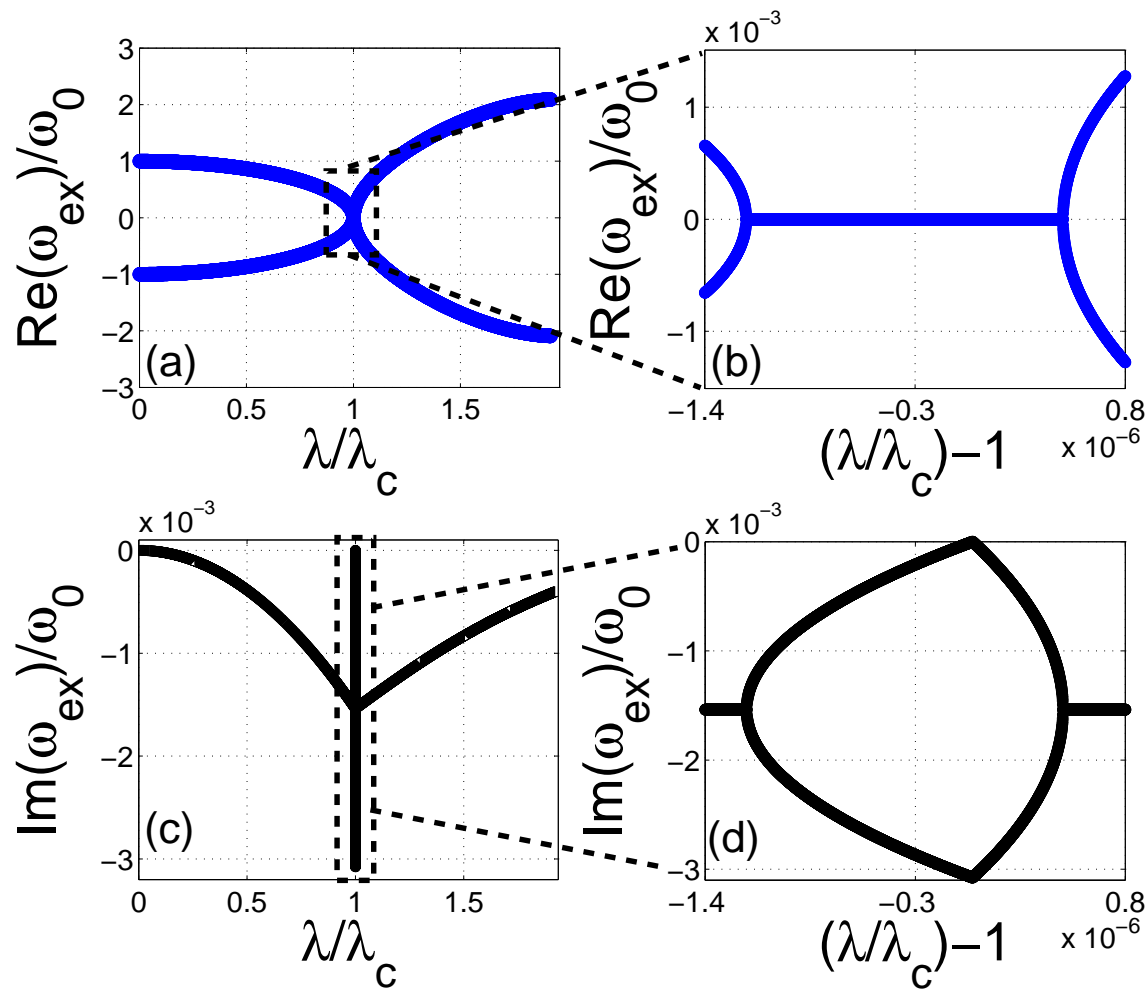}
\caption{The real and imaginary parts of the eigenfrequencies of {\it polaritonic} excitations. The figures on the right column show the indicated magnified views of real and imaginary parts around $\lambda = \lambda_c$. The parameters are $\omega/\omega_{0} = 300$, $\kappa/\omega_{0} = 200$ and for these choices of parameters $\lambda_c /\omega_{0} \approx 10.41$.}
\label{fig:fluct_eigenval}
\end{figure}

In this section, we investigate the collective light-matter excitations of the system that play a crucial role in the underlying phase transition. A detailed discussion of the excitation spectrum employing Holstein-Primakoff transformation is given in~\ref{sec:app-3}, following closely the methods of Ref.~\cite{carmichael07}. Here we present the results for the dispersive cavity regime of the self-organization problem \cite{esslinger10,nagy10} for which $\omega_0^2 \ll \omega^2 + \kappa^2$. In this regime, the real and imaginary parts of the lowest excitation eigenvalues are shown in Fig.~\ref{fig:fluct_eigenval} with $\lambda' = 0$. The real parts of the remaining two eigenvalues are very large compared to those shown in the figure due to the dispersive nature of the cavity and are not displayed. By employing perturbation theory in the small parameter $\varepsilon = \omega_0^2 / (\omega^2 + \kappa^2)$, the real and imaginary parts of the polaritonic eigenvalue up to $O(\varepsilon^2 )$ are calculated as
\be
\omega_{ex} = \omega_0 \sqrt{1 - \frac{\lambda^2}{\lambda_c^2}} \left( 1 + \frac{1}{2}\frac{\omega_0^2}{\omega^2 + \kappa^2} \frac{\lambda^2}{\lambda_c^2} \right) - i \frac{\kappa\omega_0^2}{\omega^2 + \kappa^2} \frac{\lambda^2}{\lambda_c^2} .
\label{eq:fluct_eigval}
\ee
We observe that the energy gap monotonously decreases and the lowest energy mode ``softens" as we approach the critical point from below. Note that this is the atomic excitation at $\lambda=0$ which gradually acquires a photonic component as we approach the critical point, therefore we refer to this collective excitation as polaritonic following the terminology used in Ref.~\cite{nagy08}. This is also the reason why the imaginary part of this excitation becomes larger, as in \Fig{fig:fluct_eigenval}, as the critical point is approached. We note however that there is a very narrow range around $\lambda_c$ in \Fig{fig:fluct_eigenval} where the excitation energy $\re (\omega_{ex})$ is zero and the damping $\im (\omega_{ex})$ decreases towards the critical point. In this regime of critical slowing down fluctuations are overdamped. This regime is characterized by two values $\lambda_1 < \lambda_c$ and $\lambda_2 > \lambda_c$ below and above threshold respectively. For the dispersive cavity case, these two values can be approximated by
\be
\lambda_1 \simeq \lambda_c \left[1 - \frac{\kappa^2 \omega_0^2}{(\omega^2 + \kappa^2 )^2}\right], \quad \lambda_2 \simeq \lambda_c \left[1 + \frac{1}{2} \frac{\kappa^2 \omega_0^2}{(\omega^2 + \kappa^2 )^2}\right] .
\label{eq:critical_region}
\ee
The behavior of $\omega_{ex}$ in this range is drastically different from \Eq{eq:fluct_eigval} and is given to order $( 1 - \lambda^2/\lambda_c^2)$ by
\be
\omega_{ex} \approx -i \frac{(\omega^2 + \kappa^2 )}{2\kappa}\left( 1 - \frac{\lambda^2}{\lambda_c^2} \right)
\label{eq:fluct_eigval_nearcritical}
\ee
below threshold.

The existence of this collective soft mode is significant for two reasons. Firstly, near the critical point this mode provides a fluctuation channel that drives the non-equilibrium phase transition. Secondly, due to the increasing light-like content of this channel, we get a first-hand look into the fluctuations around the critical point by monitoring photons that leak out of the cavity as pointed out in Ref.~\cite{carmichael07}.

\section{Photodetection measurements of Cavity Photons} \label{sec:measurement}

Photon flux that leaks out of the cavity below the threshold is very low. Therefore, photodetection measurements appear to be most useful for characterizing critical fluctuations. We consider below two measurement schemes, direct photon counting and second order photon correlations that can be implemented with standardly available photon counters. In \Sec{sec:modulation}, we discuss a modulation scheme which provides a direct access to the softening behavior of the polaritonic mode.

To this end, we make use of the standard input-output formalism \cite{carmichael07,milburn-book}, introducing the photonic input and output operators $\hat{a}_{in}$, $\hat{a}_{out}$, that couple to the intra-cavity fluctuation operators $\hat{c}$ and $\hat{d}$ in the following manner:

\numparts
\bea \label{eq:fluct_eom_a}
\dot{\hat{c}} = -(i\omega + \kappa) \hat{c} -i g_2 (\hat{d} + \hat{d}^{\dagger}) + \sqrt{2\kappa}\hat{a}_{in}, \\
\dot{\hat{d}} = -i\omega_0 ' \hat{d} - 2i g_1 (\hat{d} + \hat{d}^{\dagger})- i g_2 (\hat{c} + \hat{c}^{\dagger}) ,
\label{eq:fluct_eom_b}
\eea 
\endnumparts
with $\hat{a}_{in}$ being operator for the quantum noise incident on the semi-transparent cavity wall satisfying the commutations relation $[\hat{a}_{in}(t), \hat{a}_{in}^{\dagger}(t')] = \delta (t - t')$. For vacuum input, the relation $\langle \hat{a}_{in}^{\dagger}(t) \hat{a}_{in}(t') \rangle = \langle \hat{a}_{in}(t) \hat{a}_{in}(t') \rangle = 0$ holds. The cavity output can then be expressed as $\hat{a}_{out}(t) = \sqrt{2\kappa}[\hat{c}(t) + \alpha_{ss}] - \hat{a}_{in}(t)$.

\subsection{Photon counting}

The photon flux measured outside the cavity can be expressed as $I_{ss} = \langle \hat{a}_{out}^\dagger (t) \hat{a}_{out} (t) \rangle = 2\kappa \langle \hat{c}^\dagger (t) \hat{c} (t) \rangle$. For the linearized equations of motion, the photon flux was numerically found to diverge at the critical point~\cite{nagy10,carmichael07} but the precise form of scaling law was not elaborated on. We find that the intra-cavity photon number is given by (see~\ref{sec:app-4})
\be
\langle \hat{c}^\dagger (t) \hat{c} (t) \rangle_{ss}  =  \frac{\lambda^2}{2\omega\omega_0 \left[ 1 - \left( \lambda/\lambda_c \right)^2 \right]} \label{eq:phnum-neq}
\ee
displaying non-equilibrium mean field critical scaling with an exponent $\gamma_{neq} = 1$. We note that this scaling is drastically different than that of the equilibrium, ground state expectation value of the intra-cavity photon number:
\be
\langle \hat{c}^\dagger (t) \hat{c} (t) \rangle_{gs}  \approx  \frac{\lambda^2}{\omega^2 \sqrt{ 1 - \left( \lambda/\lambda_c \right)^2 }}
\ee
obtained through perturbation theory in $\varepsilon = \omega_0 / \omega$. The latter displays an exponent $\gamma_{eq} = 1/2$ as expected from standard equilibrium mean-field theory. The reason behind this is the depletion of the ground state (for $T=0$) through coupling to the photonic environment. This possibility was hinted on in Ref~\cite{nagy10}, but the authors only calculated the rate of depletion in the short-time limit i.e. in the transient regime. We find here that the depletion settles at a steady state, giving rise to an entirely different scaling law for incoherent fluctuations of intracavity photons as critical point is approached. Note that the two expressions do not even agree for $\kappa \rightarrow 0$ that enters the non-equilibrium expression through the form of $\lambda_c$. However, it should be pointed out that the limit $\kappa \rightarrow 0$ is a singular limit and has to be considered with care. The steady-state regime where this expression holds is shifted to $t \rightarrow \infty$ as $\kappa \rightarrow 0$. This time scale can be calculated from the imaginary part of the polaritonic excitation branch and is found to scale as $|\im (\omega_{ex})|^{-1} \approx \omega^3 / (4\omega_0 \lambda^2 \kappa )$ according to \Eq{eq:fluct_eigval}. For $\kappa=0$, a steady state will never be reached unless the system starts out in the ground state. What remains to be seen is whether open quantum systems may be forming an entirely new universality class. We note that the scaling behavior found in Eq.~(\ref{eq:phnum-neq}) is similar to the sub-threshold scaling of photon number in the laser~\cite{carmichael-book1} and the degenerate parametric oscillator~\cite{carmichael-book2} and reflect the fact that the damping rate of a mode of the system tends to zero at threshold, with the steady-state photon number inversely proportional to this damping rate. In contrast to these examples however, a well-defined equilibrium limit is accessible here when $\kappa \rightarrow 0$.

\subsection{Second order correlations of photons}

The second order correlation function of leaking photons $g^{(2)}(t,\tau)$ is given by
\be
g^{(2)}(t,\tau) = \frac{\langle \hat{a}_{out}^{\dagger}(t)\hat{a}_{out}^{\dagger}(t + \tau )\hat{a}_{out}(t + \tau )\hat{a}_{out}(t) \rangle}{\langle \hat{a}_{out}^{\dagger}(t) \hat{a}_{out}(t) \rangle^2}.
\label{eq:g2_output}
\ee

\begin{figure}[ht]
\centering
\subfigure[]{
\includegraphics[width=5.8cm, angle=90]{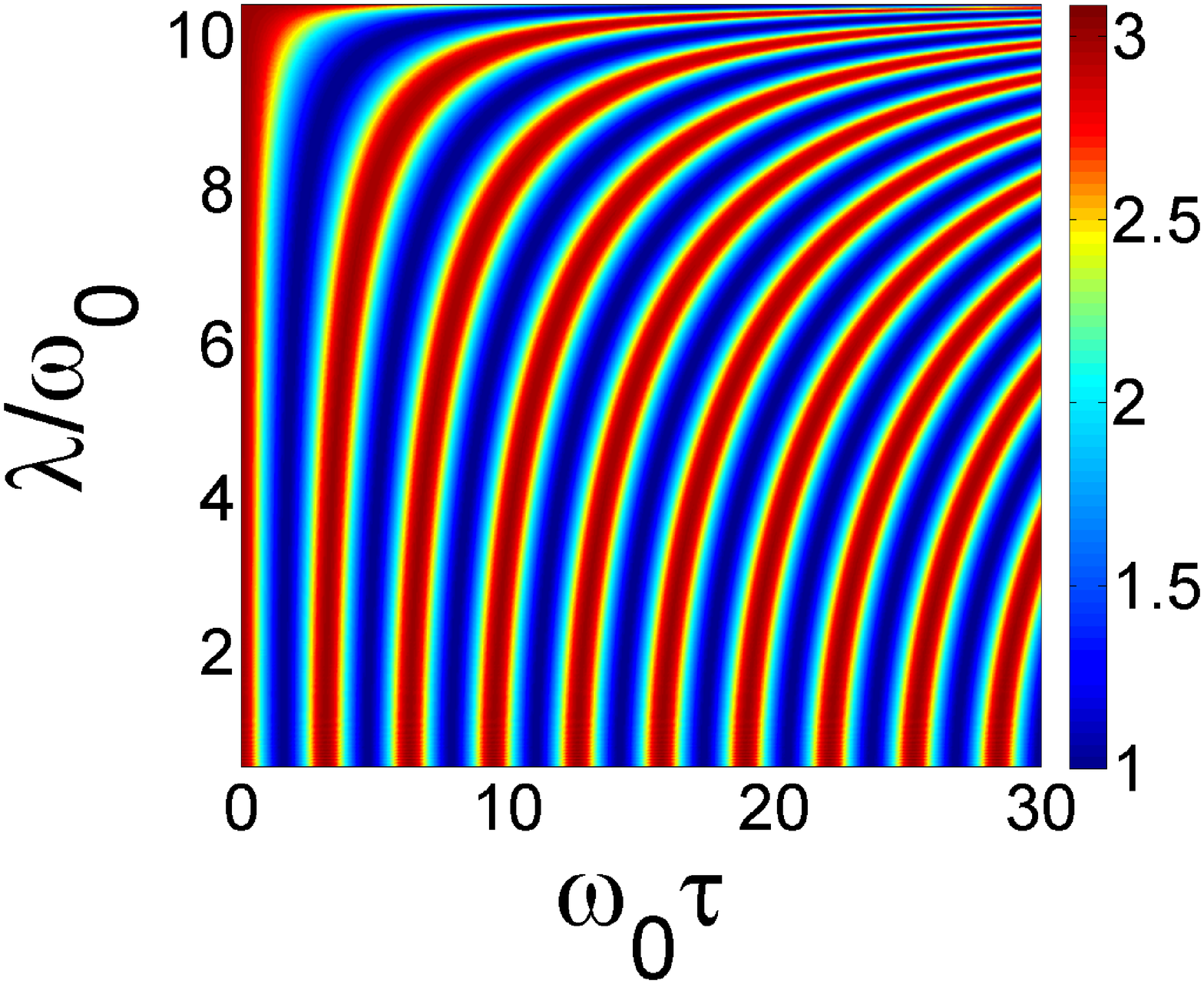}
} 
\subfigure[]{
\includegraphics[width=5.8cm, angle=90]{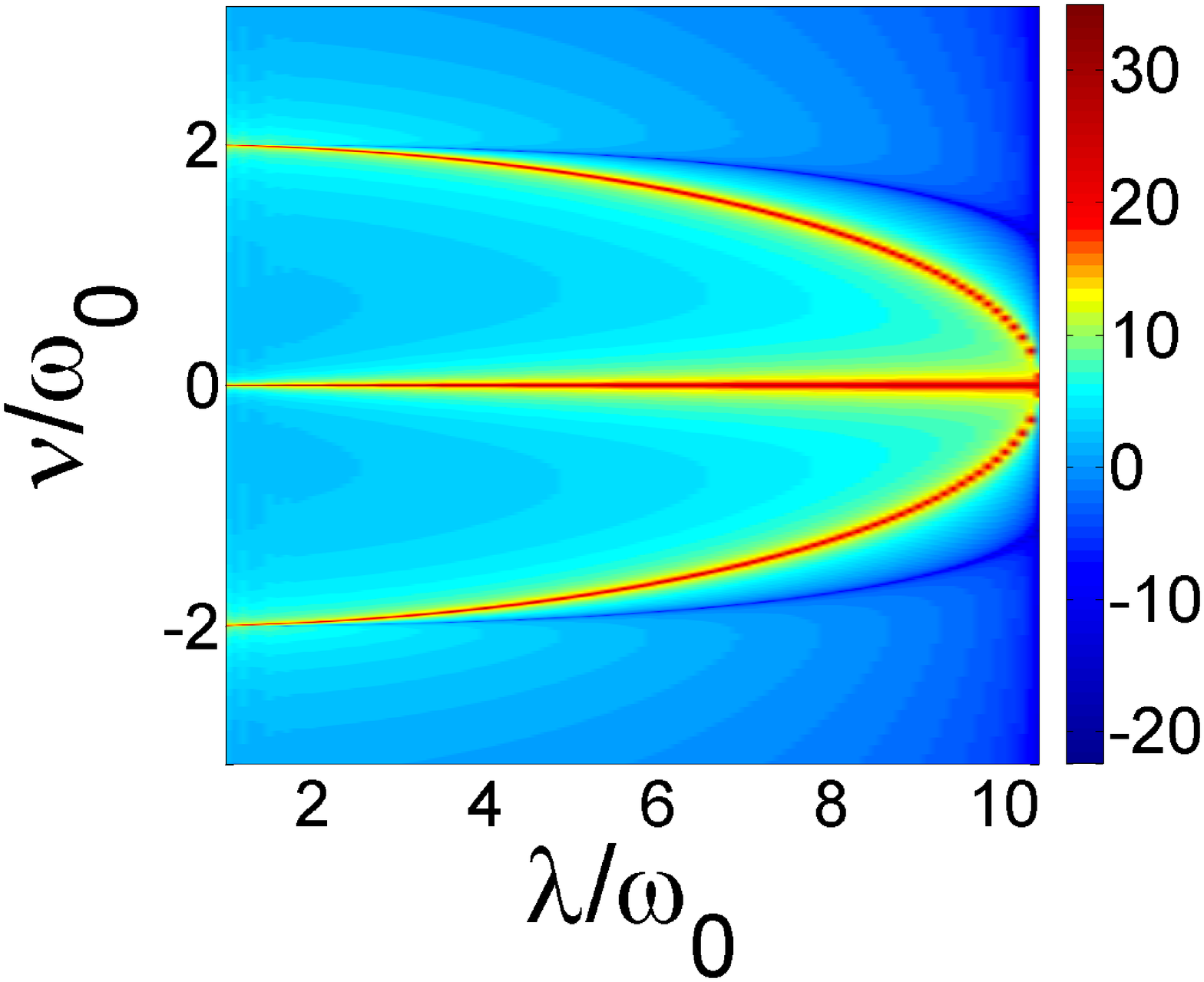}
}
\subfigure[]{
\includegraphics[width=9.5cm]{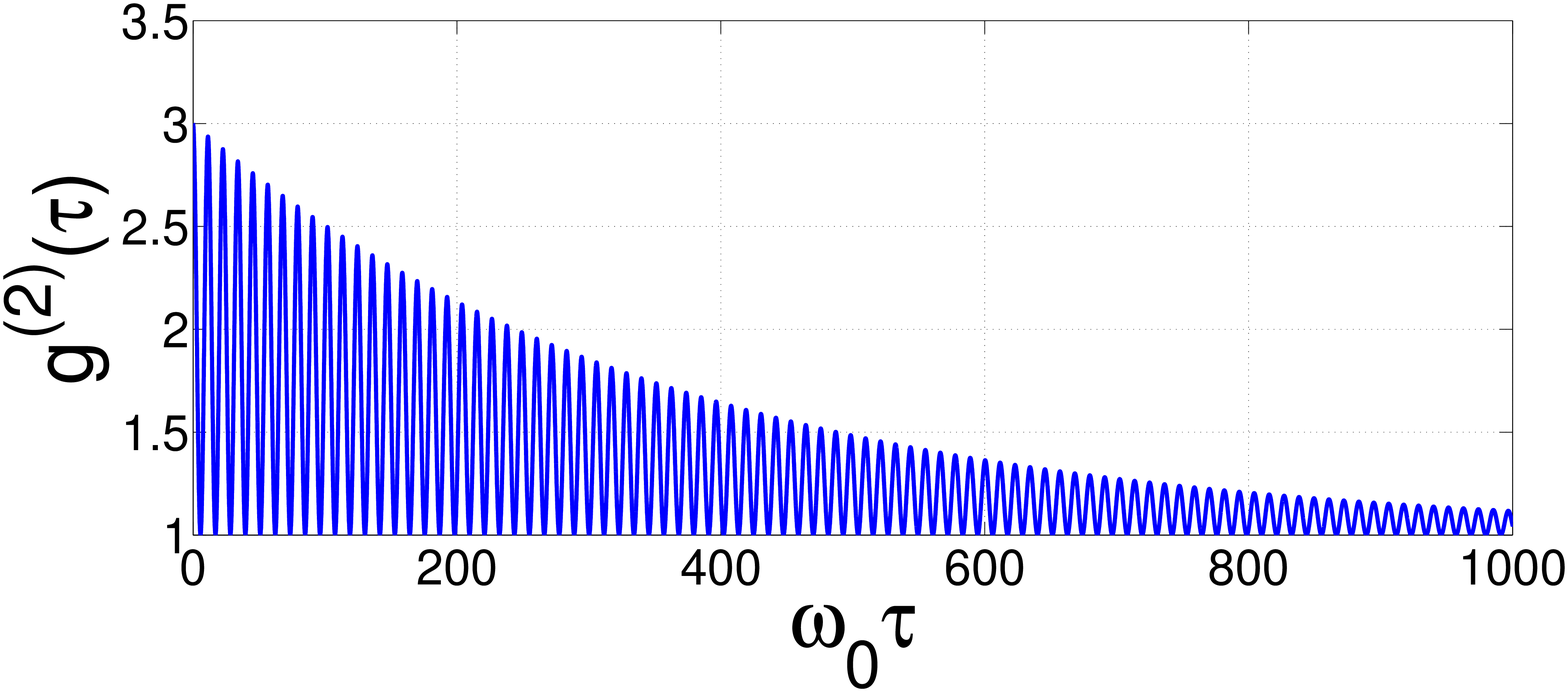}
}
\caption{(a) False color plot of the dependence of the steady-state second order correlation function $g^{(2)}_{ss}(\tau)$ on $\lambda / \omega_0$. We observe that the period of the oscillations increase with increasing $\lambda$. Note that $g^{(2)}_{ss}(\tau=0,\lambda )=3$. (b) The logarithm of the Fourier transform of $g^{(2)}_{ss}$ (color code) is plotted as a function of the scaled frequency $\nu / \omega_0$ and the scaled coupling parameter $\lambda / \omega_0$. The two peaks other than the one at $\nu = 0$ appear at $\nu = \pm2\omega_0 \sqrt{1 - (\lambda^2 / \lambda_c^2 )}$. In both figures $\lambda' = 0$. The system parameters are same as in Fig.~\ref{fig:fluct_eigenval}. (c) The long-time behavior of $g^{(2)}_{ss}(\tau)$ displaying an exponential decay of the envelope of its oscillations.}
\label{fig:g2}
\end{figure}
By using the input-output relations and the vacuum nature of the input noise, in the steady-state the expression in Eq.~(\ref{eq:g2_output}) can be simplified to (see~\ref{sec:app-5} for details)
\be
g^{(2)}(t,\tau ) = 1 + |g^{(1)}(t,\tau )|^2 
+ \frac{|\langle\hat{c}(t + \tau )\hat{c}(t) \rangle + \alpha_{ss}^2 |^2 - 2|\alpha_{ss}|^4}{(\langle\hat{c}^{\dagger}(t)\hat{c}(t) \rangle + |\alpha_{ss}|^2 )^2} ,
\label{eq:g2ttau}
\ee
where $g^{(1)}(t,\tau ) = [\langle\hat{c}^{\dagger}(t + \tau )\hat{c}(t) \rangle  + |\alpha_{ss}|^2 ] / [\langle\hat{c}^{\dagger}(t)\hat{c}(t) \rangle + |\alpha_{ss}|^2 ]$ is the first order correlation function. This steady-state form can be calculated by solving the Eqs.~(\ref{eq:fluct_eom_a}) and~(\ref{eq:fluct_eom_b}) together with their adjoints in Fourier space. Some of the results for $\lambda' = 0$ are shown in Fig.~\ref{fig:g2} for different $\lambda < \lambda_c$ values. It can be seen that $g_{ss}^{(2)}(\tau )$ displays underdamped oscillations with the oscillation period progressively increasing as critical point is approached. A Fourier analysis of the oscillations in $g_{ss}^{(2)}(\tau )$ as a function of $\lambda$ (\Fig{fig:g2}(b)) reveals that the position of the peaks $\nu_{peak}$ follow very closely the excitation frequency of the softening polaritonic mode given by \Eq{eq:fluct_eigval}. Indeed it can be shown that $\nu_{peak} (\lambda) \approx 2\re [\omega_{ex} (\lambda)]$. The width of the peaks, not resolvable for parameters chosen here, is proportional to $\im [\omega_{ex} (\lambda)]$. One can furthermore show that $g_{ss}^{(2)}(0) = 3$ by using the relation in Eq.~(\ref{eq:g2ttau}) (see~\ref{sec:app-5} for details).

\begin{figure}[ht]
\centering
\includegraphics[width=16cm]{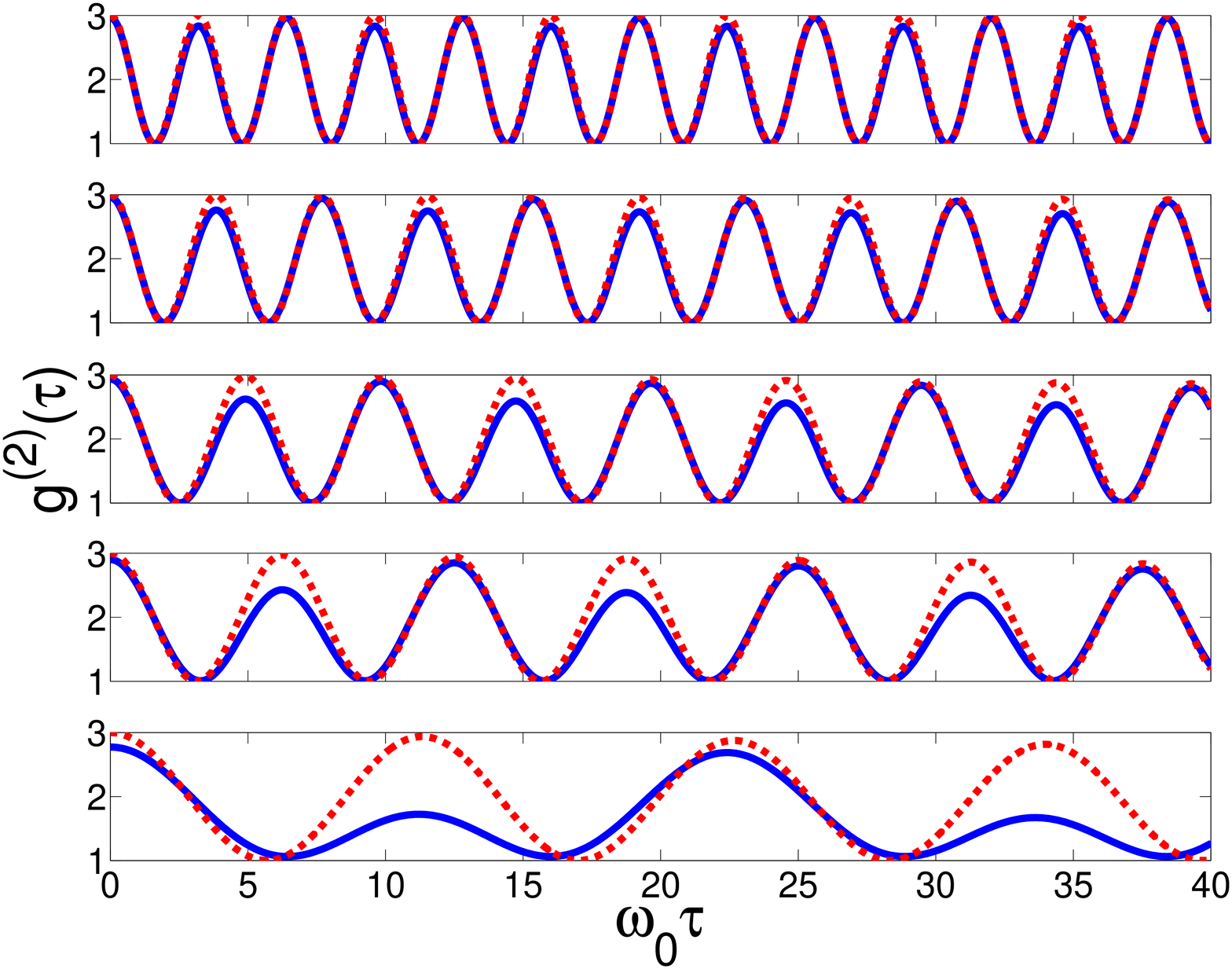}
\caption{$g^{(2)}_{ss}(\tau )$ for $\lambda' = \lambda /360$ is represented by blue solid lines and for the case $\lambda'=0$ (i.e. $\alpha_{ss} = 0$) by dashed red lines. From top to bottom,  $\lambda / \omega_0 = 2, 6, 8, 9, 10$. The asymmetry in adjacent peaks is due to the presence of nonzero mean-field $\alpha_{ss}$. The system parameters are as in Fig.~\ref{fig:fluct_eigenval}.}
\label{fig:g2_lambdap}
\end{figure}

We next calculate $g^{(2)}_{ss}(\tau )$ for nonzero values of $\lambda'$. Here, the second order correlation function displays a beating pattern in time as seen in \Fig{fig:g2_lambdap}. This is the consequence of a non-vanishing mean-field $\alpha_{ss}$ extending all the way below the threshold, providing a non-zero dc component $\alpha_{ss}$ in \Eq{eq:g2ttau}. Thus such a distinct beating pattern is a signature of the interference of a non-zero coherent cavity field and incoherent photons.

\subsection{Modulation Spectroscopy} \label{sec:modulation}

In this subsection, we analyze a modulation technique that provides an alternative access window into critical fluctuations. This technique relies on parametric resonances of the cavity-BEC system and is similar to modulation techniques applied to the analysis of ultracold atomic gases~\cite{tozzo05,esslinger04-1,esslinger04-2,esslinger05}.

While various parameters of the system may be modulated, we choose one that appears to be experimentally most straightforward: modulation of the transverse pump power. We assume a periodic modulation of the pump Rabi frequency, $\eta (t) = \eta [1 + \epsilon \cos(\nu t)]$ where $\nu$ is the modulation frequency and $\epsilon \ll 1$, resulting in the modulation of the coupling parameter $\lambda(t) = \lambda [1 + \epsilon \cos(\nu t)]$ in the Dicke model, \Eq{eq:dicke_H}. Here we analyze the case $\lambda' = 0$. The analysis below can straightforwardly be extended to non-zero $\lambda'$.

\begin{figure}[ht]
\centering
\subfigure[]{
\includegraphics[width=7.5cm]{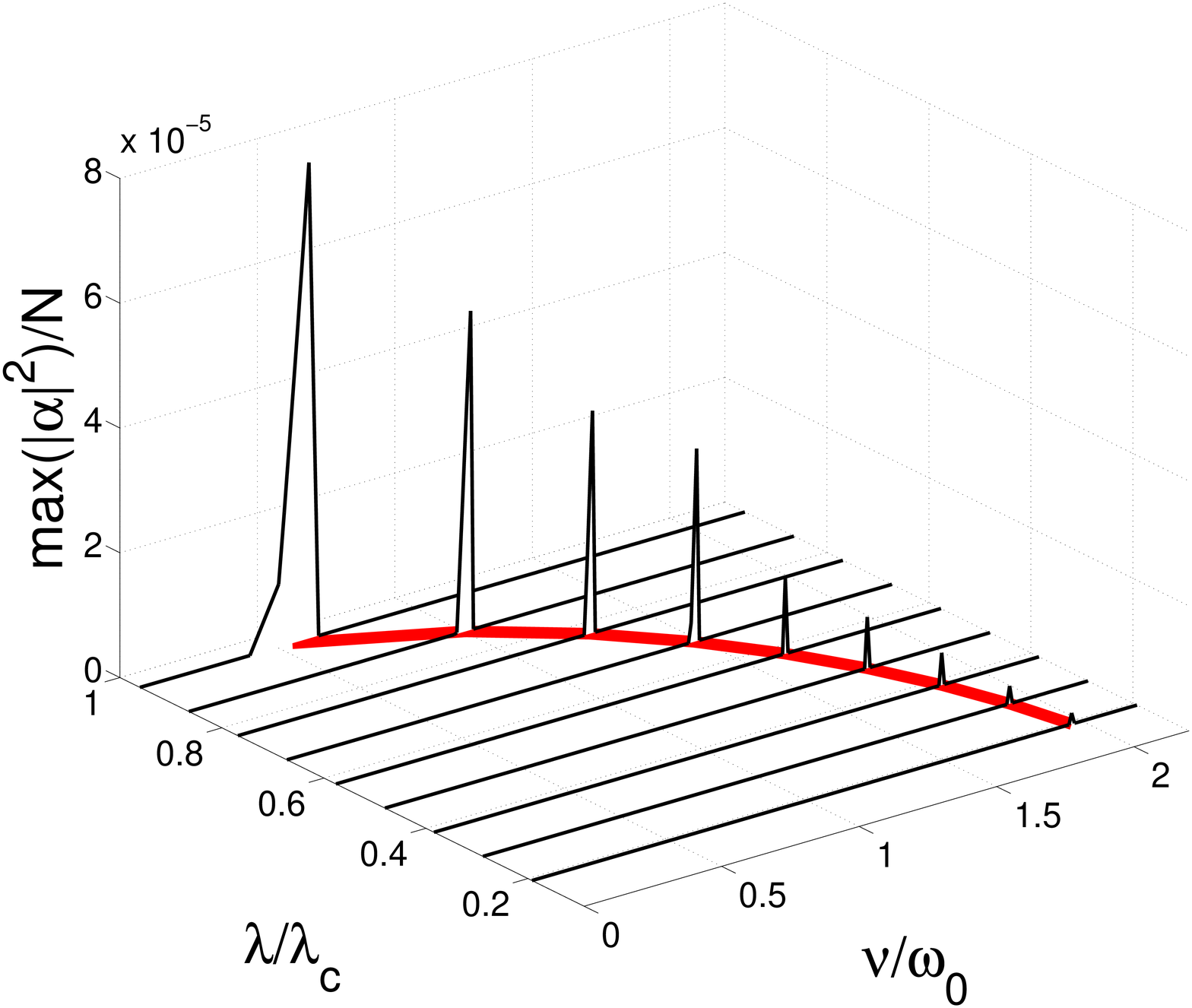}
\label{fig:modulation_alpha}
}
\subfigure[]{
\includegraphics[width=7.5cm]{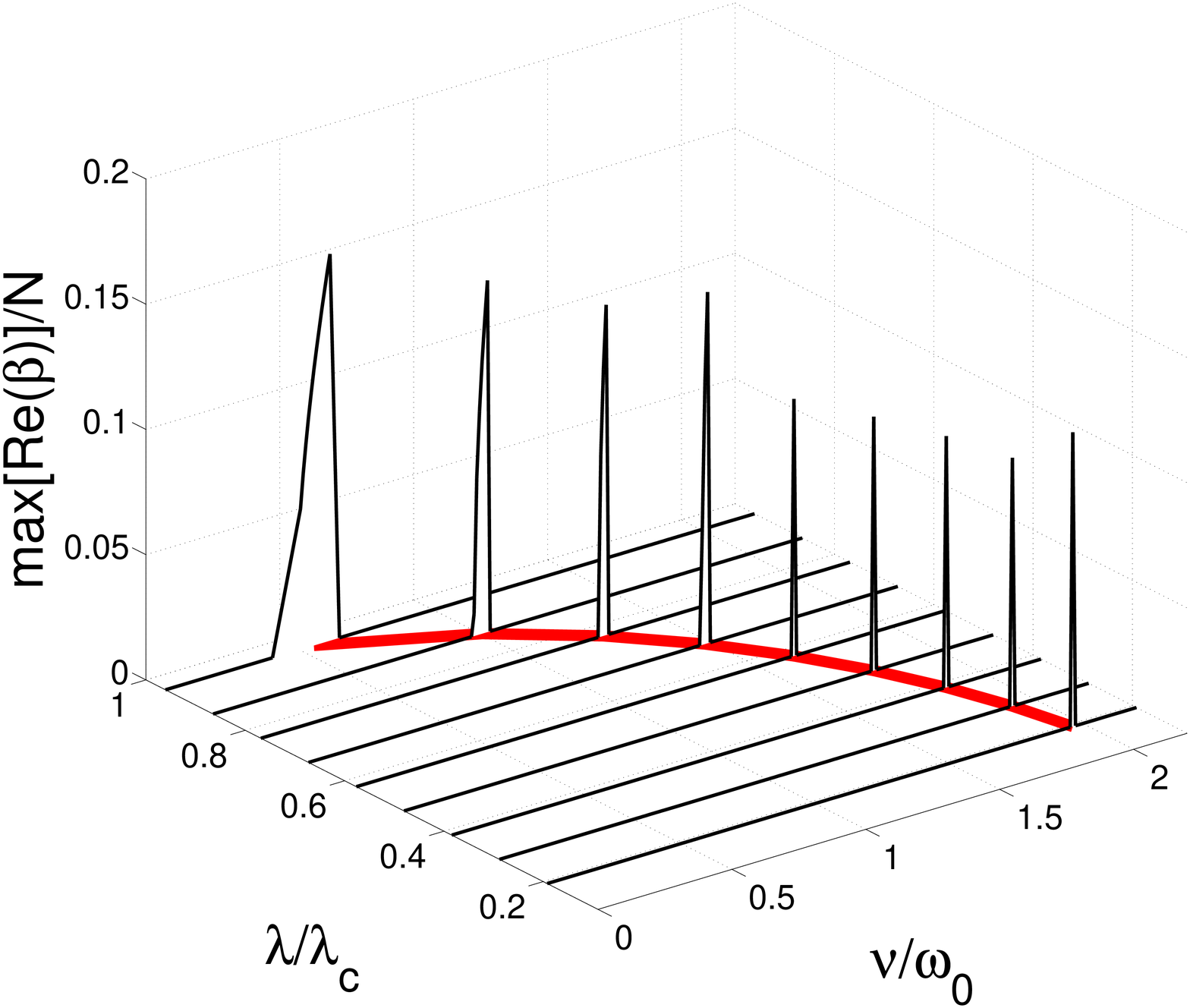}
\label{fig:modulation_beta}
}
\subfigure[]{
\includegraphics[width=8cm]{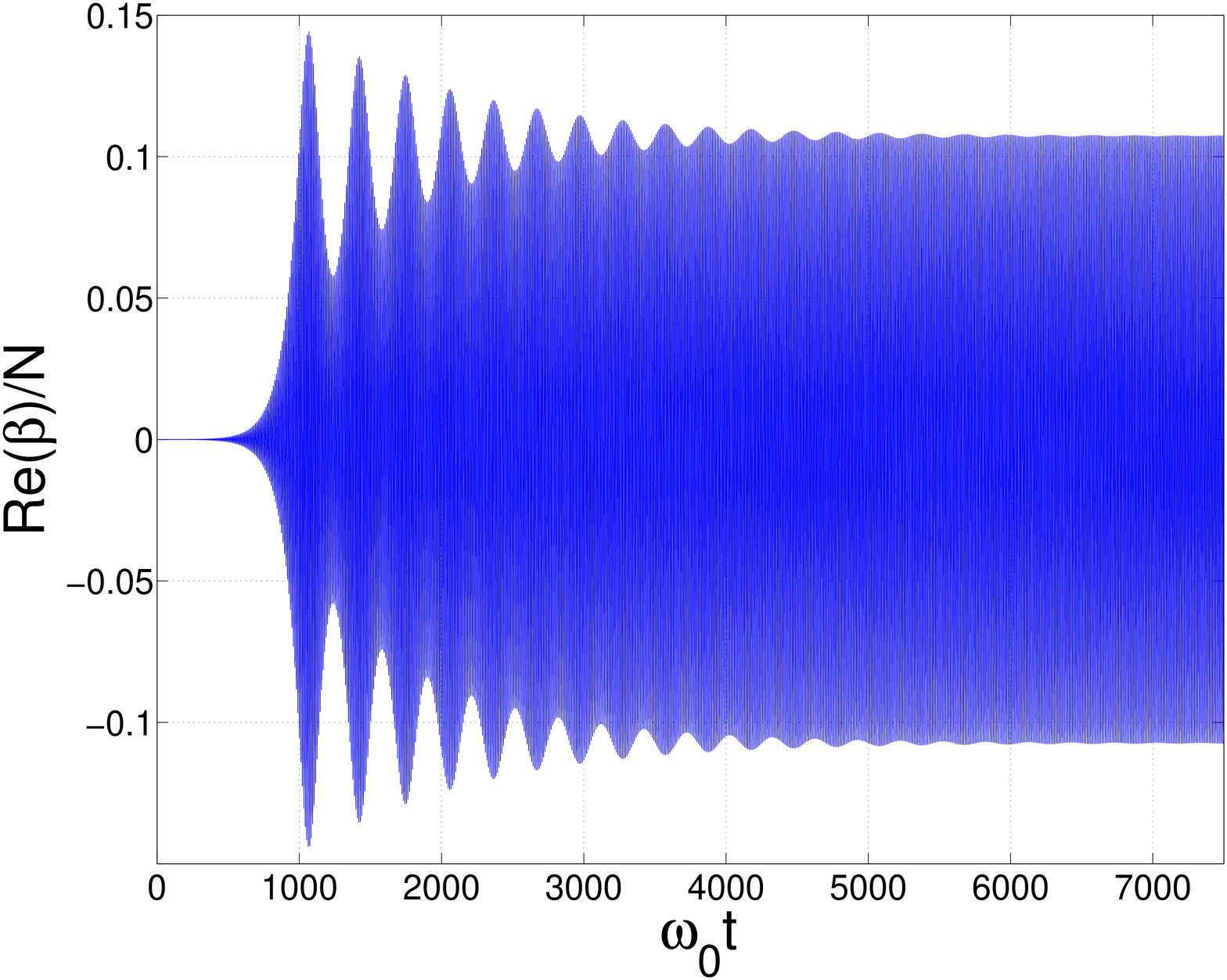}
\label{fig:dynamics_beta}
}
\caption{(a) The scaled maximum number of photons $|\alpha |^2 / N$ and (b) the scaled maximum value for $\re (\beta ) / N$, as functions of the coupling strength $\lambda / \lambda_c$ and the scaled modulation frequency $\nu / \omega_0$. The peaks occur at $\nu = 2\re (\omega_{ex})$ where $\omega_{ex}$ is given by \Eq{eq:fluct_eigval}. In both figures the red curves show the instability condition \Eq{eq:modulation_freq}. (c) The oscillations of $\re (\beta )/N$ in time are shown for $\lambda = 0.8 \lambda_c$ where the modulation frequency is chosen to be equal to resonance frequency given by $\nu = 1.2\omega_0$. For all figures here, $\epsilon = 1/50$ and system parameters are same as in Fig.~\ref{fig:fluct_eigenval}.
}
\label{fig:modulation}
\end{figure}

We assume $\kappa \gg \omega_0$, so that the cavity field dynamics adiabatically follows the atomic dynamics and the photon field $\alpha$ can be eliminated adiabatically. This results in the following equation for the atomic variable $\beta$:
\be
\dot{\beta} = -i\omega_0 \beta + 4i \lambda^2(t) \frac{\omega}{\omega^2 + \kappa^2} \left(\frac{1}{4} - |\beta |^2 \right)^{1/2} (\beta + \beta^{*}) ,
\label{eq:adielim_beta}
\ee
where we used the stable solution for $w$ which is negative. Writing $\beta = \beta_{ss} + \delta\beta (t)$, assuming small fluctuations around the steady state, we obtain the following equation for $u(t) \equiv \delta\beta (t) + \delta\beta^{*} (t)$
\be
\frac{\partial^2}{\partial\tilde{t}^2} u + \left[ A - 2\tilde{\epsilon} \cos\right(\frac{\nu}{\omega_0}\tilde{t} \left) \right] u = 0 ,
\label{eq:mathieu}
\ee
This is the well known Mathieu equation with $A = 1 - (\lambda / \lambda_c)^2$, $\tilde{\epsilon} = (\lambda / \lambda_c)^2 \epsilon$ and $\tilde{t} = \omega_0 t$. According to the Floquet theorem, the solutions to this equation have the form $u(\tilde{t}) = \exp (\mu \tilde{t}) \phi(\tilde{t})$ where $\phi(\tilde{t})$ is a periodic function with period $2\pi \omega_0 /\nu$~\cite{orszag-book}. One can readily see that the solutions $u(\tilde{t})$ are unstable if $\re (\mu ) > 0$. We performed a perturbative stability analysis to determine the stability boundary for the solutions~\cite{orszag-book}. We find that the instability appears when the condition
\be
\nu = 2\omega_0 \sqrt{1 - \left( \frac{\lambda}{\lambda_c}\right)^2} ,
\label{eq:modulation_freq}
\ee
is fulfilled. This leading order expression is precisely twice the excitation frequency of Eq.~(\ref{eq:fluct_eigval}). Finally, we check our result by solving the system of coupled nonlinear equations~(\ref{eq:alpha_eom}) and~(\ref{eq:beta_eom}) in real time. In \Fig{fig:modulation}, we plot for each modulation frequency $\nu$ the maximum number of photons after the oscillation has stabilized, starting with very small initial ($\alpha$, $\beta$). Thus when the frequency of modulation is chosen correctly, there will be a substantial photon flux at the cavity face, even below threshold. We should also note that there is a broadening in resonance modulation frequencies as the critical point is approached (see \Fig{fig:modulation}), and this is consistent with the behavior of the imaginary part of the eigenmode in \Fig{fig:fluct_eigenval}.

\section{Conclusion} \label{sec:conclusion}

In this work, we discussed the excitations of an optically driven atomic condensate coupled to a single mode of a high-finesse cavity that displays a self-organization transition as a function of the driving strength. Taking into account the finite-size of the system, we showed that the coupled BEC-cavity system can be mapped into an open-system realization of the Dicke model with a symmetry-breaking field.

The zero-field phase transition is driven by softening of polaritonic excitations, which provides access to the internal dynamics of the coupled system close to the criticality. We discuss a number of photodetection-based techniques for the photons that leak out of the cavity mirrors and relate it to intra-cavity critical dynamics.

We find that the intra-cavity photon number, which can be measured by a photodetector outside the cavity, displays a different scaling law for the open system than the closed system with perfectly reflecting mirrors. In the latter case, the dynamics conserves total excitation number and the photon number is calculated in the ground state.

We next discuss the second order correlation function for photons $g^{(2)}_{ss}(\tau )$ in the steady state, via coincidence measurement of photons leaking out of the cavity walls. We show that $g^{(2)}_{ss}(\tau )$ displays damped oscillations with a period that increases progressively as the critical point is approached, signalling the critical slowing down of the coupled intra-cavity dynamics. We relate the spectral content of the time-series of $g^{(2)}_{ss}(\tau )$ to the complex frequency of the softening polaritonic mode. We also show that trap misalignment can lead to a background coherent cavity field and that its signature is a characteristic beating pattern in $g^{(2)}_{ss}(\tau )$.

Finally, we discuss a modulation scheme that directly captures the softening of the polaritonic mode. This is done by introducing a parametric instability in the system through the periodic modulation of the drive Rabi frequency in time. We show that modulation at twice the polaritonic mode frequency results in a measurable photon flux outside the cavity below the threshold.

\ack
The authors thank K. Baumann, F. Brennecke, T. Donner, R. Mottl and T. Esslinger for fruitful discussions and insight into experimental data, closely related to the measurement schemes presented here. We thank D. A. Huse, A. Pal, S. Schmidt, M. Schiro, S. Shinohara for stimulating discussions. We also thank P. Domokos for bringing to our attention their recent preprint~\cite{nagy11} closely related to the work reported in this paper. H.E.T. acknowledges support from Swiss NSF under Grant No. PP00P2-123519/1. \"O.E.M. acknowledges support by TUBITAK for the Project No. 109T267.

\appendix
\section{}  \label{sec:app-1}

Here we discuss some details of the derivation of Hamiltonian in Eq.~(\ref{eq:manybody_H}). We assume that the pump-cavity detuning $|\Delta_c| = |\omega_p - \omega_c| \sim \kappa$ so that the cavity can be quasi-resonantly excited by the scattered pump photons from atoms. We assume a high-finesse cavity where the pump photons are dominantly scattered into a single mode; coupling to multiple modes is possible in the bad cavity limit or for resonators with degenerate modes \cite{gopalakrishnan-pra10}. Simultaneously, the laser is red-detuned far from an internal atomic transition at $\omega_a$, so that $|\Delta_a| = |\omega_p - \omega_a| \gg \gamma_{a}$, where $\gamma_{a}$ is the atomic linewidth. This ensures that the atoms are predominantly in their ground states during the excitation process, suppressing spontaneous emission and giving rise to an optical potential for the motional degrees of freedom of atoms $|\hat{E}^+(\bm{x})|^2/\Delta_a$. Here $\hat{E}^+(\bm{x}) = g_0 \varphi_c (\bm{x})  \aop +  \Omega_p \varphi_p(\bm{x})$ is the positive rotating component of the electric field felt by an atom of the cloud at position $\bm{x}$, due to the interference of the cavity field (photon creation operator $\hat{a}$, atom-field coupling $g_0$, mode function $\varphi_c (\bm{x})$) and a coherent laser field (with Rabi frequency $\Omega_p$ and standing wave pattern $\varphi_p(\bm{x})$). The Hamiltonian under these approximations is given by
\be
\hat{H} = -\hbar\Delta_c \hat{a}^{\dagger} \hat{a} + \int d\bm{x} \, \hat{\Psi}^{\dagger}(\bm{x}) \left[ \frac{-\hbar^2}{2m}\nabla^2 + \hbar\frac{|\hat{E}^+(\bm{x})|^2}{\Delta_a} + V(\bm{x}) \right] \hat{\Psi}(\bm{x})  .
\ee
We consider a situation where the BEC is trapped by an additional external trapping potential $V(\bm{x})$ and that this is deep in the radial direction confining the cloud along the cavity axis ($x$). With the additional assumption that the driving laser beam is broad, the problem can be reduced to an effective one-dimensional problem with $\varphi_p(\bm{x}) \approx const.$ (the constant to be absorbed into $\Omega_p$) and $\varphi_c(\bm{x}) \approx \varphi_c(x)$~\cite{ritsch08}. We will assume that the cavity mode function is given by $\varphi_c(x) = \frac{1}{\sqrt{L}} \sin (Gx)$ where $G = 2\pi/\lambda_{cav} \approx 2\pi/\lambda_p$ with $\lambda_{cav}$ and $\lambda_p$ being the cavity mode and pump wavelengths respectively. The final Hamiltonian is then given by
\bea
\hat{H} &=& -\hbar\Delta_c \hat{a}^{\dagger} \hat{a} + \int dx \, \hat{\Psi}^{\dagger}(x) \left[ \frac{-\hbar^2}{2m}\frac{d^2}{dx^2} + V(\bm{x}) \right. \nonumber \\
&+& \left. \hbar U_0 |\varphi_c (x)|^2 \hat{a}^{\dagger}\hat{a} + \hbar\eta \varphi_c (x) (\hat{a}^{\dagger} + \hat{a}) \right] \hat{\Psi}(x),
\label{eq:manybody_H_app}
\eea
where $U_0 = g_0^2/\Delta_a$ and $\eta = \Omega_p g_0/\Delta_a$. We have subtracted the energy provided by the constant potential $\hbar\Omega_p^2/\Delta_a$.

\section{} \label{sec:app-2}
In this section, we present the steady-state analysis of the semi-classical equations of motion for the non-equilibrium Dicke model by taking into account the effects of the symmetry-breaking field term $\lambda^\prime$. The discussion in this appendix follows closely that of Ref.~\cite{carmichael07}. We will however be interested in the far detuned regime $\omega \gg \omega_{0}$ relevant to experiments reported in Ref.~\cite{esslinger10}.

For the mean field expressions $\alpha = \langle \hat{a}\rangle $, $\beta = \langle \hat{J}_{-}\rangle$ and $w=\langle \hat{J}_{z}\rangle$, the Heisenberg equations of motion for the system Hamiltonian~(\ref{eq:dicke_H}) can be written as

\numparts
\bea
\label{eq:meanfield_eom}
\dot{\alpha} = -(\kappa + i\omega)\alpha - i\frac{\lambda}{\sqrt{N}}(\beta + \beta^{*}) - i\frac{\lambda^\prime}{\sqrt{N}}\left(\frac{N}{2} - w\right) , \label{eq:alpha_eom}\\
\dot{\beta} = -i\omega_0 \beta + 2i \frac{\lambda}{\sqrt{N}}(\alpha + \alpha^{*})w + i\frac{\lambda^\prime}{\sqrt{N}}\beta (\alpha + \alpha^{*}) , \label{eq:beta_eom}\\
\dot{w} = i\frac{\lambda}{\sqrt{N}}(\alpha + \alpha^{*})(\beta - \beta^{*}) , \label{eq:w_eom}
\eea
\endnumparts

\begin{figure}[htb]
\centering
\subfigure[]{
\includegraphics[width=7cm]{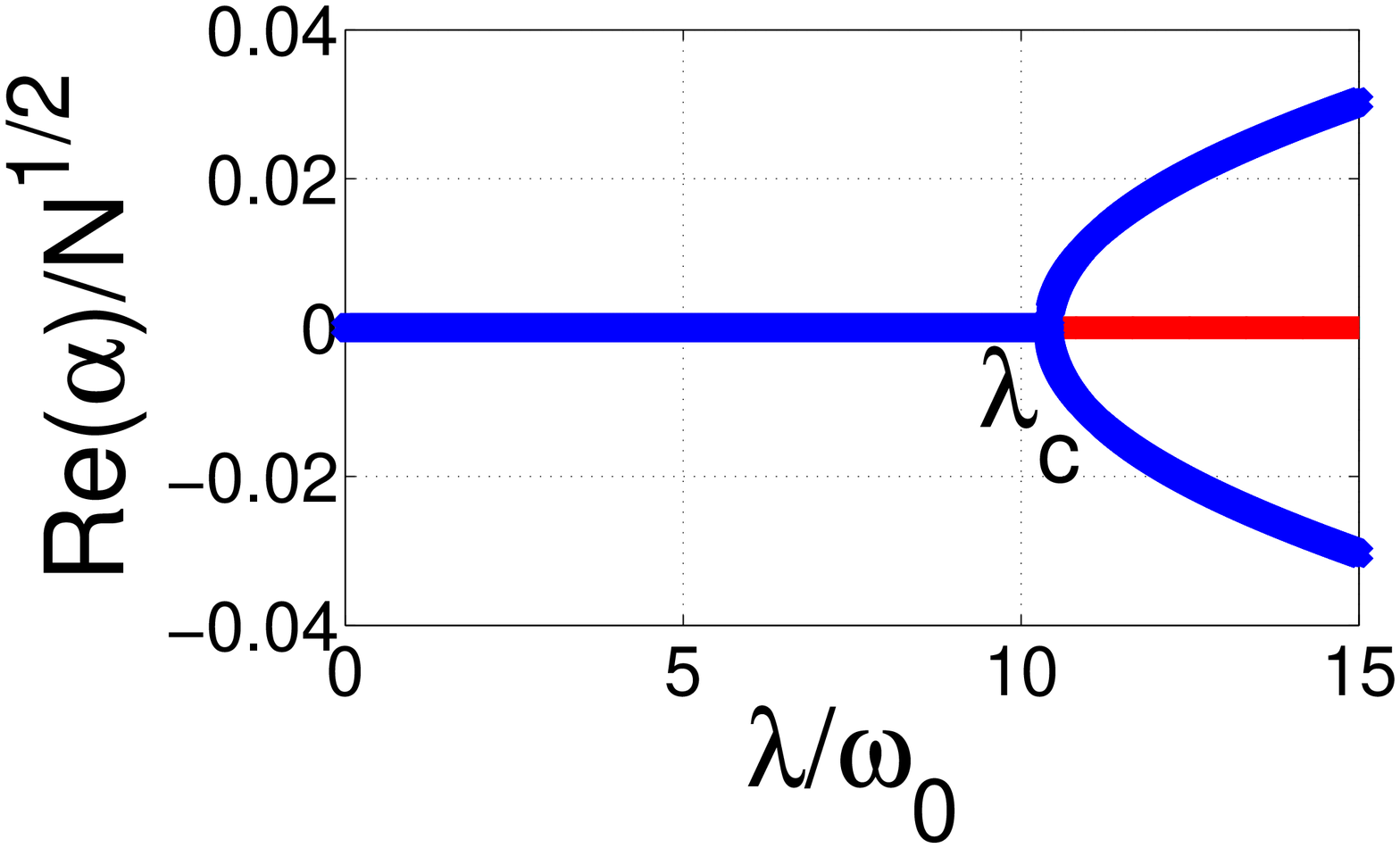}
\label{fig:alp_lpzero}
}
\subfigure[]{
\includegraphics[width=7cm]{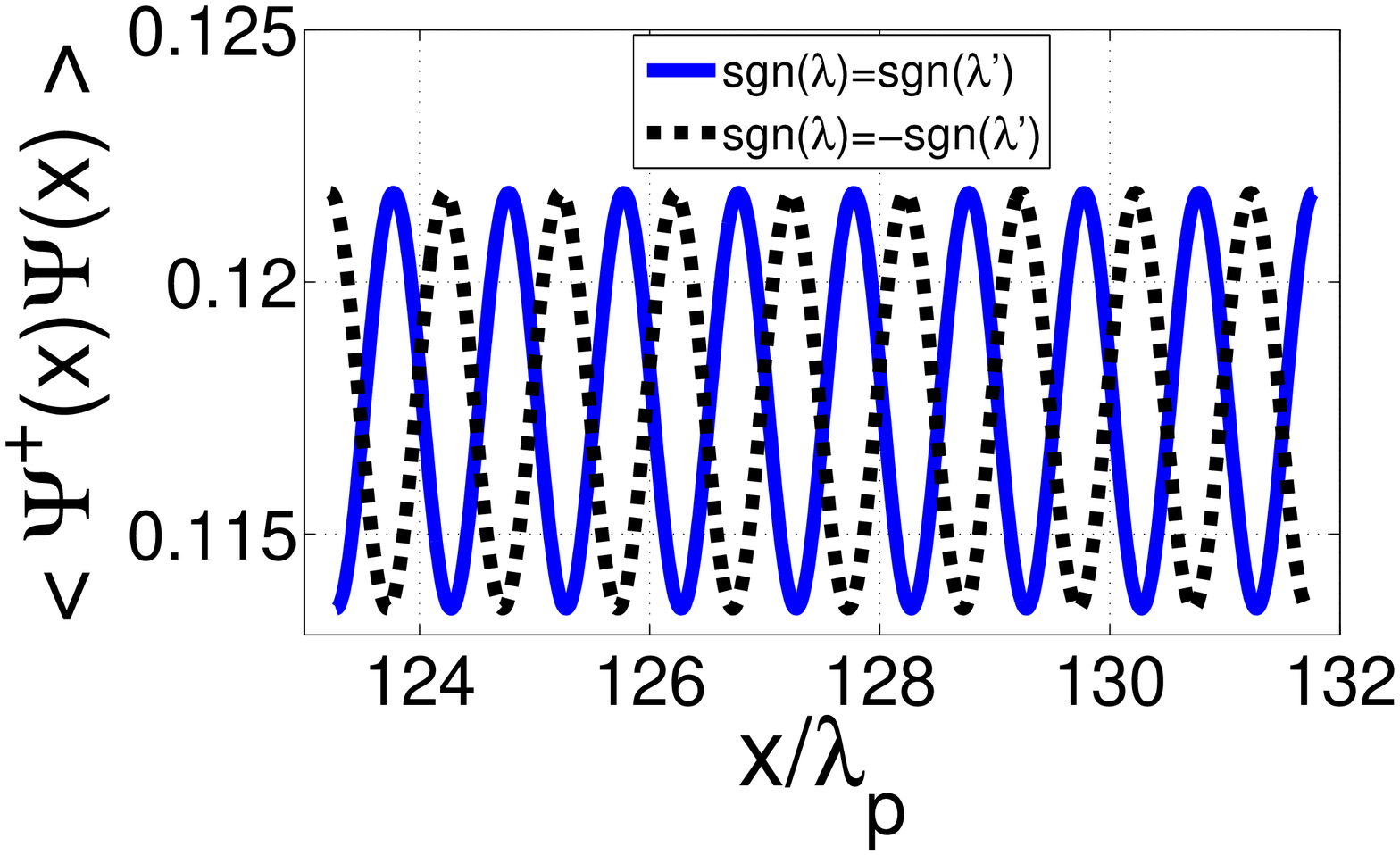}
\label{fig:density}
}
\subfigure[]{
\includegraphics[width=7cm]{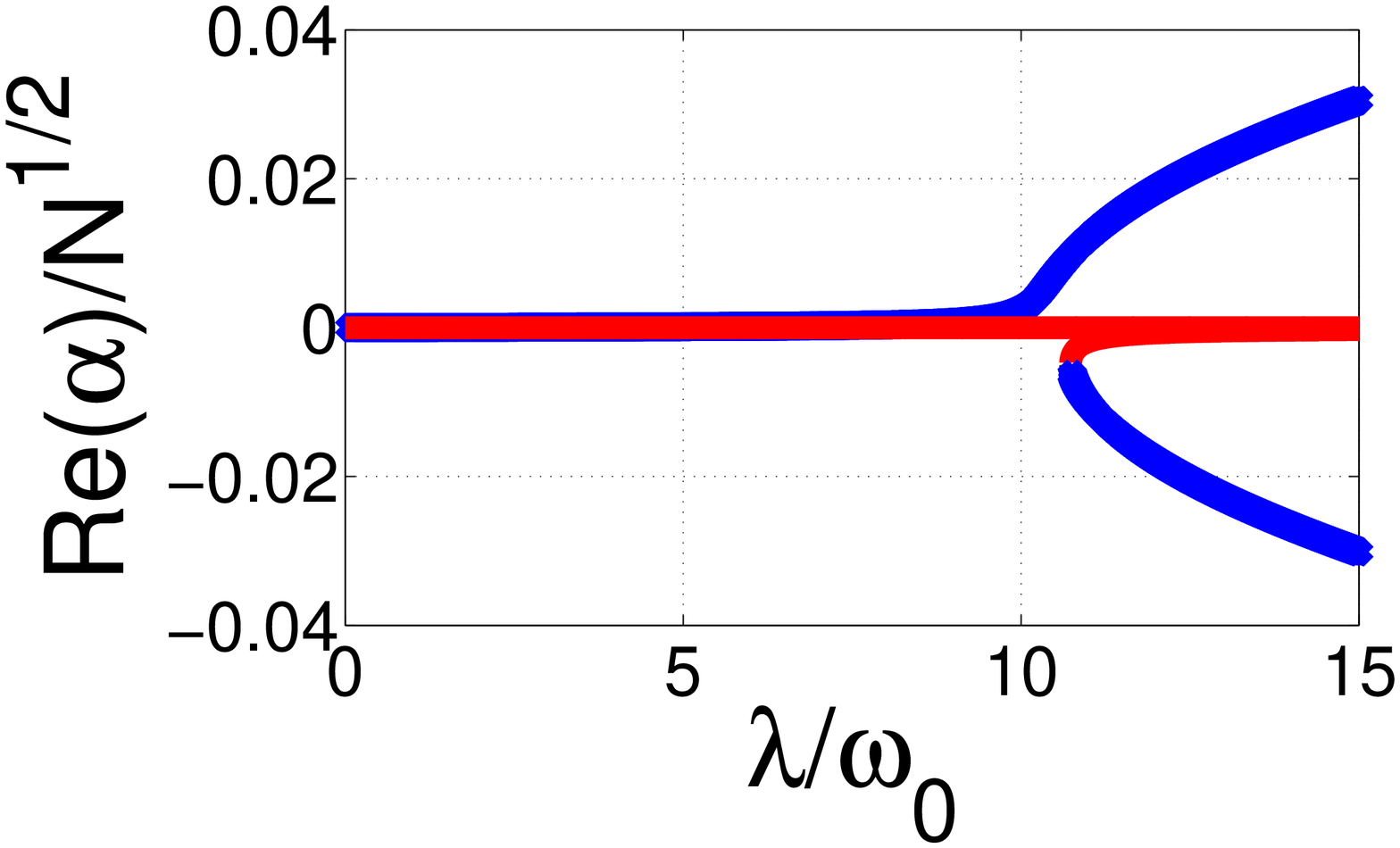}
\label{fig:alp_lppos}
}
\subfigure[]{
\includegraphics[width=7cm]{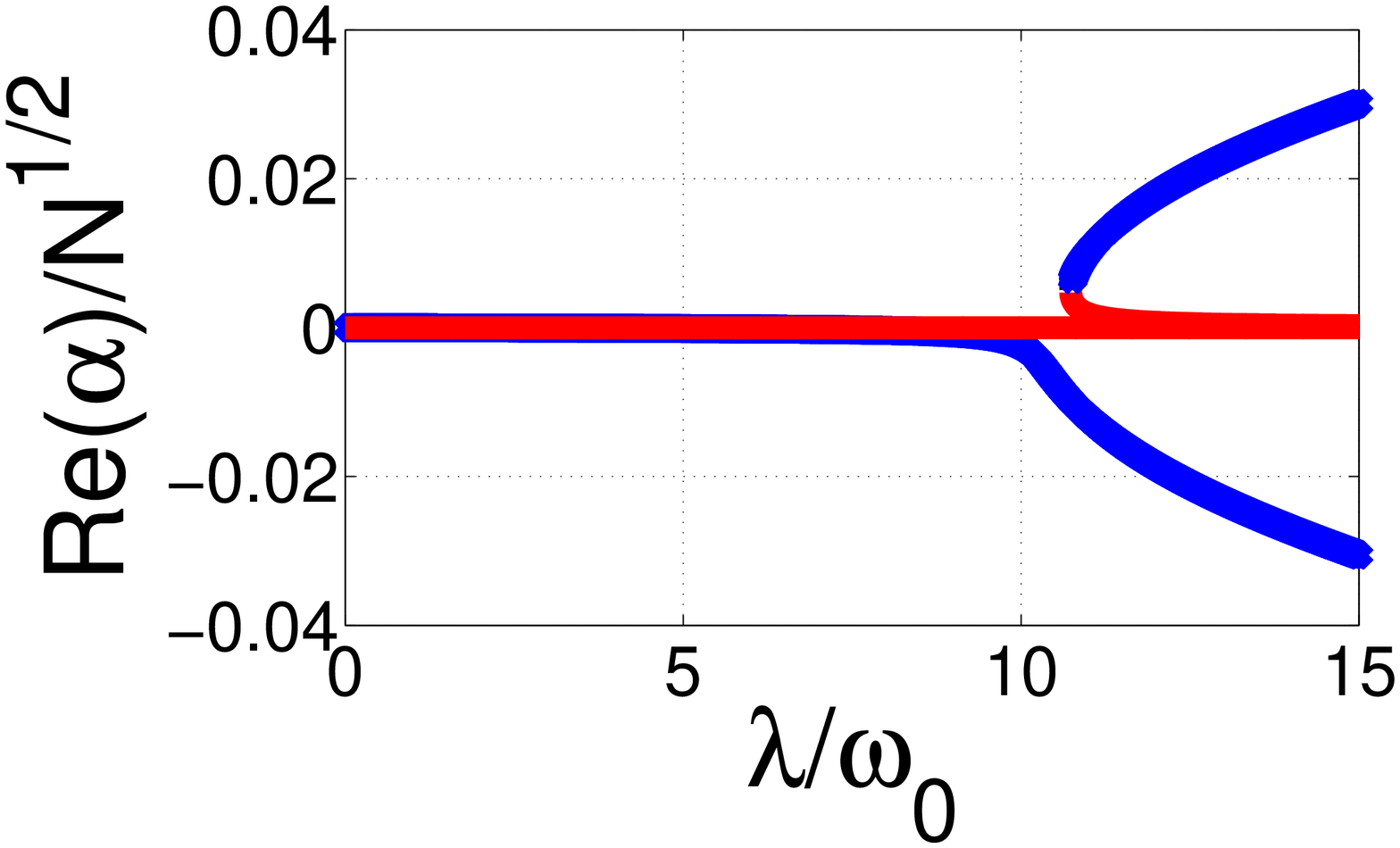}
\label{fig:alp_lpneg}
}
\caption{(a) The real part of $\alpha$ for $\lambda' = 0$. Below threshold the stable solution is $\alpha = 0$ and above threshold there are two stable solutions with a phase difference $\pi$. (b) The steady state atomic density profiles as functions of scaled position $x/\lambda_p$ (where $\lambda_p$ is the pump wavelength) for $\lambda' < 0$ and $\lambda' > 0$ with $\lambda = 9\omega_0$ and $\lambda' / \lambda \approx \pm120$. (c) Dependence of $\alpha$ on the coupling parameter for the trap displaced to the right, which corresponds to $sgn(\lambda ) = sgn(\lambda' )$. (d) Same as (c) except that the displacement of the trap is to the left so that $sgn(\lambda ) = -sgn(\lambda' )$. In (a), (c) and (d) blue (red) lines represent stable (unstable) solutions. The system parameters are same as in Fig.~\ref{fig:fluct_eigenval}.}
\label{fig:symmetry_breaking}
\end{figure}

These equations have to be solved with the constraint that the pseudo angular momentum $|\beta |^2 + w^2 = N^2 /4$ is conserved. Analytical solutions can be found for this set of nonlinear equations in the steady-state when $\lambda^\prime = 0$~\cite{carmichael07}. In that case, the steady state solution displays a bifurcation point at $\lambda = \lambda_c$. While $\alpha_{ss}=\beta_{ss}=0$ is the trivial solution for all values of $\lambda$, it's only stable for $\lambda < \lambda_c$. For $\lambda > \lambda_c$, this solution becomes unstable and two new sets of stable solutions appear given by~\cite{carmichael07}
\numparts
\bea
\label{eq:ss_abovethreshold}
\alpha_{ss} = \pm \sqrt{N} \frac{\lambda}{\omega - i\kappa} \sqrt{1 - \frac{\lambda_c^4}{\lambda^4}} , \label{eq:alpha_ss} \\
\beta_{ss} = \mp \frac{N}{2}\sqrt{1 - \frac{\lambda_c^4}{\lambda
^4}} , \label{eq:beta_ss}
\eea
\endnumparts
The behavior of these solutions for $\alpha_{ss}$ are shown in \Fig{fig:symmetry_breaking}(a). For $\lambda' \neq 0$, the bifurcation point (and the threshold) disappears and the stable solution becomes nonzero for all $\lambda > 0$ values (\Fig{fig:symmetry_breaking}(c,d)). We do not plot $\beta$, it displays a similar behavior to $\alpha$, with its sign opposite to that of $\alpha$. We note that the spatial structure of the self-organized state, as shown in \Fig{fig:symmetry_breaking}(b), depends on the sign of $\lambda'$.

\section{} \label{sec:app-3}
In this Appendix, we discuss using Holstein-Primakoff representation in order to study the collective excitations of the system. The Holstein-Primakoff transformation can be applied by expressing the atomic spin operators in terms of bosonic mode operators $\hat{b}$ and $\hat{b}^{\dagger}$ such as $\hat{J}_{+} = \hat{b}^{\dagger}\sqrt{N - \hat{b}^{\dagger}\hat{b}}$, $\hat{J}_{-} = \hat{J}_{+}^{\dagger}$ and $\hat{J}_{z} = \hat{b}^{\dagger}\hat{b} - N/2$~\cite{brandes-pre03,carmichael07}. Substituting these expressions into the Dicke Hamiltonian~(\ref{eq:dicke_H}) and expanding in the limit $N\gg 1$, we get the Hamiltonian governing the fluctuations around the steady state semiclassical solutions $\alpha_{ss}$ and $\beta_{ss}$
\be
\hat{H}_{HP}/\hbar = \omega \hat{c}^{\dagger} \hat{c} + \omega_0 ' \hat{d}^{\dagger} \hat{d} + g_1 (\hat{d}^{\dagger} + \hat{d})^2 + g_2 (\hat{c}^{\dagger} + \hat{c}) (\hat{d}^{\dagger} + \hat{d})
\label{eq:dicke_HP}
\ee
where $\hat{c}$ and $\hat{d}$ are the photonic and atomic fluctuation operators respectively, $\hat{a} = \alpha_{ss} + \hat{c}$ and $\hat{b} = \beta_{ss}/\sqrt{N} + \hat{d}$, and

\numparts
\bea
\label{eq:new_couplings}
\omega_0 ' = \omega_0 - 2\lambda \frac{\tilde{\beta}_{ss}}{\sqrt{1 - \tilde{\beta}_{ss}^2 }} \re (\tilde{\alpha}_{ss} ) , \\
g_1 = -\lambda \frac{\tilde{\beta}_{ss} (2 - \tilde{\beta}_{ss}^2 )}{2(1 - \tilde{\beta}_{ss}^2 )^{3/2}} \re (\tilde{\alpha}_{ss} )  , \\
g_2 = \lambda \frac{1 - 2\tilde{\beta}_{ss}^2}{\sqrt{1 - \tilde{\beta}_{ss}^ 2}} - \lambda' \tilde{\beta}_{ss} .
\eea
\endnumparts

Here we introduced the scaled variables $\tilde{\alpha}_{ss} = \alpha_{ss}/\sqrt{N}$, $\tilde{\beta}_{ss} = \beta_{ss}/N$. One should note that the steady-state values $\alpha_{ss}$ and $\beta_{ss}$ are $\lambda'$-dependent. The quadratic Hamiltonian above leads to linear equations of motion $\mathbf{\dot{h}} = \mathbf{M}\mathbf{h}$ for fluctuations $\mathbf{h} = (\hat{c} , \hat{c}^{\dagger}, \hat{d}, \hat{d}^{\dagger})$. This was discussed in the resonant case $\omega = \omega_0$ and $\lambda' = 0$ in Ref.~\cite{carmichael07}. 

\section{} \label{sec:app-4}
The expression in Eq.~(\ref{eq:phnum-neq}) for the number of photons and $g^{(2)}_{ss}(0)$ below threshold are calculated by using the quantum regression theorem to write the equations of motion for two-operator product averages. The system of linear equations becomes
\bea
\label{eq:lin-sys}
\frac{d}{dt}\langle \hat{c} \hat{c} \rangle = -(2 i \omega + 2\kappa) \langle \hat{c} \hat{c} \rangle - 2i\lambda (\langle \hat{c} \hat{d} \rangle + \langle \hat{c} \hat{d}^{\dagger} \rangle ) , \nonumber \\
\frac{d}{dt}\langle \hat{c}^{\dagger} \hat{c} \rangle = -2\kappa \langle \hat{c}^{\dagger} \hat{c} \rangle - i\lambda (\langle \hat{c}^{\dagger} \hat{d}^{\dagger} \rangle + \langle \hat{c}^{\dagger} \hat{d} \rangle - \langle \hat{c} \hat{d}^{\dagger} \rangle - \langle \hat{c} \hat{d} \rangle ) , \nonumber  \\
\frac{d}{dt}\langle \hat{d}^{\dagger} \hat{d}^{\dagger} \rangle = 2 i \omega _0 \langle \hat{d}^{\dagger} \hat{d}^{\dagger} \rangle + 2i\lambda (\langle \hat{c}^{\dagger} \hat{d}^{\dagger} \rangle + \langle \hat{c} \hat{d}^{\dagger} \rangle ) , \nonumber \\
\frac{d}{dt}\langle \hat{d}^{\dagger} \hat{d} \rangle = i\lambda (\langle \hat{c}^{\dagger} \hat{d} \rangle + \langle \hat{c} \hat{d} \rangle - \langle \hat{c} \hat{d}^{\dagger} \rangle - \langle \hat{c}^{\dagger} \hat{d}^{\dagger} \rangle ) , \nonumber  \\
\frac{d}{dt}\langle \hat{c} \hat{d} \rangle = -(i\omega_0 + i\omega + \kappa) \langle \hat{c} \hat{d} \rangle - i\lambda (\langle \hat{c} \hat{c} \rangle + \langle \hat{d} \hat{d} \rangle + \langle \hat{c}^{\dagger} \hat{c} \rangle + \langle \hat{d}^{\dagger} \hat{d} \rangle + 1) , \nonumber \\
\frac{d}{dt}\langle \hat{c}^{\dagger} \hat{d} \rangle = (-i\omega_0 + i\omega - \kappa) \langle \hat{c}^{\dagger} \hat{d} \rangle + i\lambda (-\langle \hat{c}^{\dagger} \hat{c}^{\dagger} \rangle + \langle \hat{d} \hat{d} \rangle - \langle \hat{c}^{\dagger} \hat{c} \rangle + \langle \hat{d}^{\dagger} \hat{d} \rangle ) , 
\eea
where we use the relations $\langle\hat{a}_{in}(t)\hat{d}(t) \rangle = \langle\hat{d}(t)\hat{a}_{in}(t) \rangle = \langle\hat{a}_{in}(t)\hat{d}^{\dagger}(t)\rangle = \langle\hat{d}^{\dagger}(t)\hat{a}_{in}(t)\rangle = 0$, which are valid for the zero temperature photonic bath outside the cavity. For the steady-state, we set all the time derivatives to zero and solve the linear system. One outcome of this calculation is the photon number expression $\langle \hat{c}^{\dagger}(t) \hat{c}(t) \rangle_{ss}$ given in Eq.~(\ref{eq:phnum-neq}). We also use the solutions obtained here in the calculation of Eq.~(\ref{eq:g2ttau}) for $\tau=0$, as expounded in Appendix E.

\section{} \label{sec:app-5}

This Appendix details the derivation of the expression Eq.~(\ref{eq:g2ttau}) from the general expression Eq.~(\ref{eq:g2_output}). Using the input-output relation $\hat{a}_{out}(t) = \sqrt{2\kappa}[\hat{c}(t) + \alpha_{ss}] - \hat{a}_{in}(t)$ in Eq.~(\ref{eq:g2_output}), all the correlators that have $\hat{a}_{in}$ operator to the right in the averages vanish due to the fact that we consider zero temperature bath. The same conclusion is true for all the correlators that have $\hat{a}_{in}^{\dagger}$ to the left. In addition to this, using the relation $[\hat{O}(t),\hat{a}_{in}(t')] = 0$ for $t < t'$ and $\hat{O}(t)$ being any system operator~\cite{milburn-book}, one can write
\be
g^{(2)}(t,\tau) = \frac{\langle [\hat{c}^{\dagger}(t) + \alpha_{ss}^{*}] [\hat{c}^{\dagger}(t + \tau ) + \alpha_{ss}^{*}] [\hat{c}(t + \tau ) + \alpha_{ss}] [\hat{c}(t) + \alpha_{ss}] \rangle}{\langle [\hat{c}^{\dagger}(t) + \alpha_{ss}^{*}] [\hat{c}(t) + \alpha_{ss}] \rangle^2} .
\label{eq:g2_system}
\ee
Since the Holstein-Primakoff Hamiltonian is quadratic and the dynamics is described by linear quantum Langevin equations Eqs.~(\ref{eq:fluct_eom_a}) and (\ref{eq:fluct_eom_b}), the three operator products can be decoupled as
\bea
\langle \hat{O}_1 \hat{O}_2 \hat{O}_3 \rangle &=& \langle \hat{O}_1 \rangle \langle \hat{O}_2 \hat{O}_3 \rangle + \langle \hat{O}_2 \rangle \langle \hat{O}_1 \hat{O}_3 \rangle + \langle \hat{O}_3 \rangle \langle \hat{O}_1 \hat{O}_2 \rangle \nonumber \\
&-& 2 \langle \hat{O}_1 \rangle \langle \hat{O}_2 \rangle \langle \hat{O}_3 \rangle .
\label{eq:3_decoupling}
\eea
Here, the operators can be at different times in general.  Single fluctuation operator averages $\langle \hat{O}_i \rangle$ vanish by the way we define fluctuation operators, hence all three operator averages vanish. Once all three operator averages are shown to be zero, four operator averages decouple as
\bea
\langle \hat{O}_1 \hat{O}_2 \hat{O}_3 \hat{O}_4 \rangle &=& \langle \hat{O}_1 \hat{O}_2 \rangle \langle \hat{O}_3 \hat{O}_4 \rangle + \langle \hat{O}_1 \hat{O}_3 \rangle \langle \hat{O}_2 \hat{O}_4 \rangle + \langle \hat{O}_1 \hat{O}_4 \rangle \langle \hat{O}_2 \hat{O}_3 \rangle .
\label{eq:4_decoupling}
\eea
Using Eq.~(\ref{eq:3_decoupling}) and setting $\langle \hat{O}_i \rangle=0$, the nonzero terms of $g^{(2)}(t,\tau)$ in Eq.~(\ref{eq:g2_system}) are
\bea
g^{(2)}(t,\tau) = \frac{|\alpha_{ss}|^4 + \langle \hat{c}^{\dagger}(t) \hat{c}^{\dagger}(t + \tau ) \hat{c}(t + \tau ) \hat{c}(t) \rangle}{[\langle \hat{c}^{\dagger}(t) \hat{c}(t) \rangle + |\alpha_{ss}|^2 ]^2} \nonumber \\
+ \frac{|\alpha_{ss}|^2 [\langle \hat{c}^{\dagger}(t) \hat{c}(t) \rangle + \langle \hat{c}^{\dagger}(t + \tau ) \hat{c}(t + \tau )\rangle + \langle \hat{c}^{\dagger}(t) \hat{c}(t + \tau ) \rangle + \langle \hat{c}^{\dagger}(t + \tau ) \hat{c}(t) \rangle ]}{[\langle \hat{c}^{\dagger}(t) \hat{c}(t) \rangle + |\alpha_{ss}|^2 ]^2} \nonumber \\
+ \frac{\alpha_{ss}^2 \langle \hat{c}^{\dagger}(t) \hat{c}^{\dagger}(t + \tau )\rangle + \alpha_{ss}^{*2} \langle \hat{c}(t + \tau ) \hat{c}(t)\rangle}{[\langle \hat{c}^{\dagger}(t) \hat{c}(t) \rangle + |\alpha_{ss}|^2 ]^2} .
\label{eq:g2_system_nonzero}
\eea
Using Eq.~(\ref{eq:4_decoupling}), the numerator of the first term on the right hand side of Eq.~(\ref{eq:g2_system_nonzero}) decouples as
\bea
\langle \hat{c}^{\dagger}(t) \hat{c}^{\dagger}(t + \tau ) \hat{c}(t + \tau ) \hat{c}(t) \rangle &=& |\langle \hat{c}(t + \tau ) \hat{c}(t) \rangle |^2 + |\langle \hat{c}^{\dagger}(t + \tau ) \hat{c}(t) \rangle |^2 \nonumber \\
&+& \langle \hat{c}^{\dagger}(t) \hat{c}(t) \rangle^2 ,
\label{eq:g2_4_decoupling}
\eea
where we use the fact that in the steady-state $\langle \hat{c}^{\dagger}(t + \tau ) \hat{c}(t + \tau ) \rangle = \langle \hat{c}^{\dagger}(t) \hat{c}(t) \rangle$. Also, the last term on the right hand side of Eq.~(\ref{eq:g2_system_nonzero}) can be written as
\bea
\frac{\alpha_{ss}^2 \langle \hat{c}^{\dagger}(t) \hat{c}^{\dagger}(t + \tau )\rangle + \alpha_{ss}^{*2} \langle \hat{c}(t + \tau ) \hat{c}(t)\rangle}{[\langle \hat{c}^{\dagger}(t) \hat{c}(t) \rangle + |\alpha_{ss}|^2 ]^2} &=& \frac{|\langle \hat{c}(t + \tau ) \hat{c}(t) \rangle + \alpha_{ss}^2 |^2 }{[\langle \hat{c}^{\dagger}(t) \hat{c}(t) \rangle + |\alpha_{ss}|^2 ]^2} \nonumber \\
&-& \frac{|\alpha_{ss}|^4 + |\langle \hat{c}(t + \tau ) \hat{c}(t) \rangle|^2}{[\langle \hat{c}^{\dagger}(t) \hat{c}(t) \rangle + |\alpha_{ss}|^2 ]^2} .
\label{eq:g2_anomalous}
\eea
If we now use the Eqs.~(\ref{eq:g2_4_decoupling}) and~(\ref{eq:g2_anomalous}) together with the expression for the first order correlation function $g^{(1)}(t,\tau) = [\langle\hat{c}^{\dagger}(t + \tau )\hat{c}(t) \rangle  + |\alpha_{ss}|^2 ] / [\langle\hat{c}^{\dagger}(t)\hat{c}(t) \rangle + |\alpha_{ss}|^2 ]$, Eq.~(\ref{eq:g2_system}) reduces to the expression in Eq.~(\ref{eq:g2ttau}). 

We note that the validity of the decoupling rules of Eq.~(\ref{eq:3_decoupling}-\ref{eq:4_decoupling}) can be easily verified by calculating multi-operator product averages in Fourier space, as in Ref.~\cite{carmichael07}. For instance, Eq.~(\ref{eq:g2ttau}) can be obtained by writing the intracavity photon operator $\hat c$ in terms of the input noise operator $\hat a_{in}$ in Fourier space and using the commutation relations for $\hat a_{in}$ and $\hat a_{in}^{\dagger}$. The form of Eq.~(\ref{eq:g2ttau}) has the virtue of explicitly singling out the contribution of anomalous averages to $g^{(2)}(t,\tau )$.

If we take delay time $\tau = 0$, below threshold, the second order correlation function in Eq.~(\ref{eq:g2ttau}) takes the form
\be
g_{ss}^{(2)}(0) = 2 + \frac{|\langle \hat c (t) \hat c (t) \rangle_{ss}|^2}{\langle \hat c^{\dagger}(t) \hat c (t) \rangle_{ss}^2}
\label{eq:g2_0}
\ee
The solution of the system of equations Eq.~(\ref{eq:lin-sys}) in the steady-state yieds
\be
\langle \hat c (t) \hat c (t)\rangle_{ss} = \frac{\lambda^2 }{2\omega\omega_0 \left(\omega^2 + \kappa^2 \right) \left[1 - \left(\lambda /\lambda_c \right)^2 \right]} \left[ \left(\omega^2 - \kappa^2 \right) + 2i\omega\kappa \right].
\label{eq:aa_exp}
\ee
Comparing this result with the expression in Eq.~(\ref{eq:phnum-neq}) for steady-state photon number below threshold, we find the relationship $\langle \hat c^{\dagger} (t) \hat c (t)\rangle_{ss} = |\langle \hat c (t) \hat c (t)\rangle_{ss}|$. This, together with Eq.~(\ref{eq:g2_0}) yields our final result $g^{(2)}_{ss}(0) = 3$.

\section*{References}
\bibliographystyle{iopart-num}
\bibliography{dicke_refs}     

\end{document}